\documentclass[sigconf]{acmart}

\usepackage{macros}

\usepackage{cuted}
\usepackage{pifont}
\usepackage{capt-of}
\usepackage{subcaption}
\usepackage{makecell}
\usepackage{csquotes}

\usepackage{csquotes}
\usepackage{amsmath}

\usepackage{soul} 
\usepackage{sparklines}

\definecolor{sparkspikecolor}{named}{blue}

\AtBeginDocument{%
  \providecommand\BibTeX{{%
    \normalfont B\kern-0.5em{\scshape i\kern-0.25em b}\kern-0.8em\TeX}}}

\setcopyright{rightsretained}

\begin{document}

\title[Storytelling in an Introductory Creative Coding Class]{Exploring Individual and Collaborative Storytelling \\in an Introductory Creative Coding Class} 

\author{Sangho Suh}
\affiliation{\institution{University of Waterloo}}
\email{sangho.suh@uwaterloo.ca}

\author{Ken Jen Lee}
\affiliation{\institution{University of Waterloo}}
\email{kenjen.lee@uwaterloo.ca}

\author{Celine Latulipe}
\affiliation{\institution{University of Manitoba}}
\email{celine.latulipe@umanitoba.ca}

\author{Jian Zhao}
\affiliation{\institution{University of Waterloo}}
\email{jian.zhao@uwaterloo.ca}

\author{Edith Law}
\affiliation{\institution{University of Waterloo}}
\email{edith.law@uwaterloo.ca}

\renewcommand{\shortauthors}{Suh et al.}

\begin{abstract}
Teaching programming through storytelling is a popular pedagogical approach and an active area of research. However, most previous work in this area focused on K-12 students using block-based programming. Little, if any, work has examined the approach with university students using text-based programming. This experience report fills this gap. Specifically, we report our experience administering three storytelling assignments---two individual and one collaborative---in an introductory computer science class with 49 undergraduate students using \textit{p5.js}, a text-based programming library for creative coding. Our work contributes an understanding of students' experiences with the three authoring processes and a set of recommendations to improve the administration of and experience with individual and collaborative storytelling with text-based programming.
\end{abstract}




\begin{teaserfigure}
\centering
\includegraphics[trim=0cm 0cm 0cm 0.05cm, clip=true, width=0.95\textwidth]{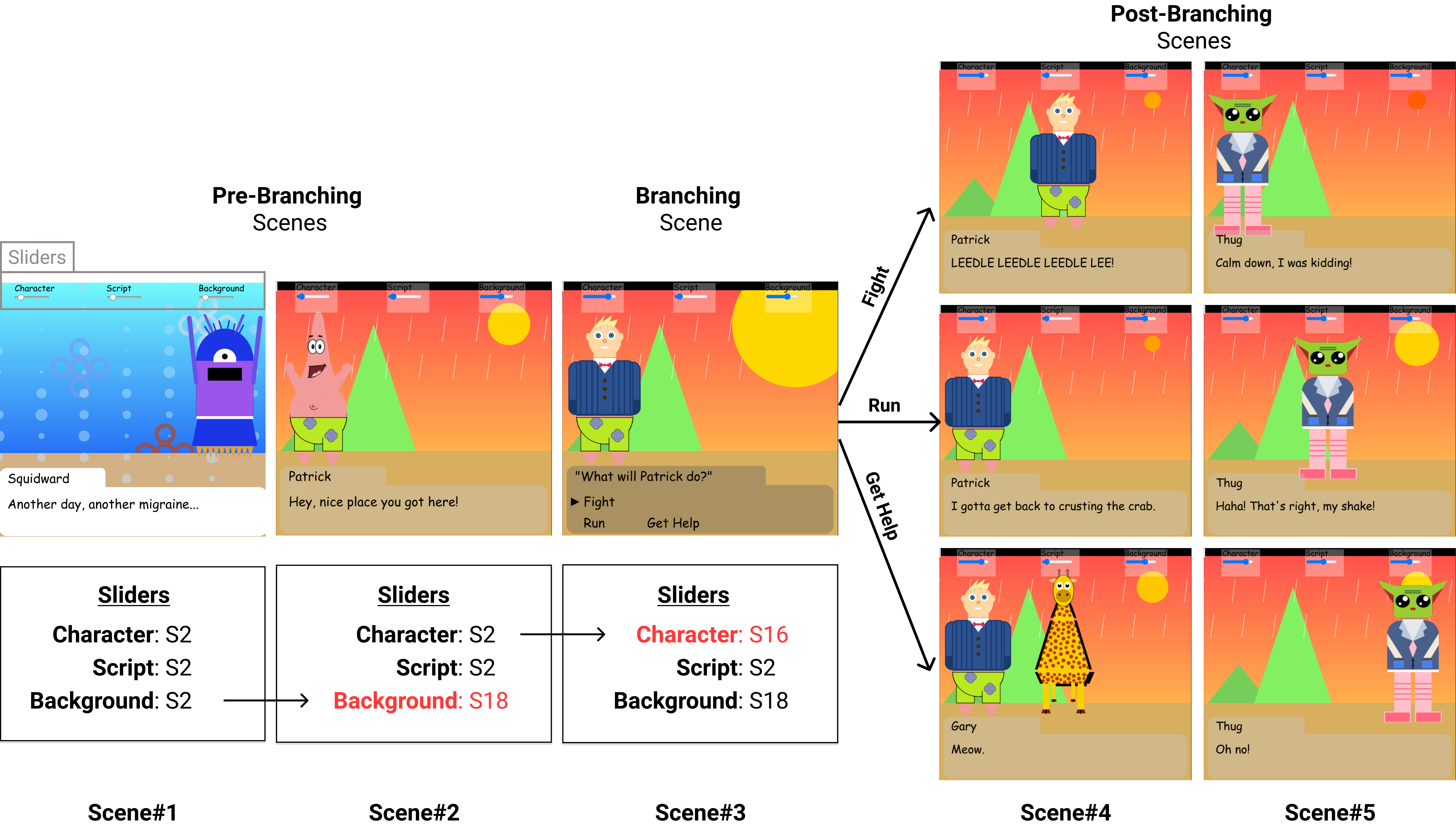}
\caption{An example of combining and remixing students' interactive narratives from our collective storytelling assignment (A6). Here, sliders are used to change background (scene\#2) and characters (scene\#3).}
\label{fig:crowdsourced_narratives}
\end{teaserfigure}

\maketitle
\section{Introduction}


Creative coding (also known as media computing~\cite{guzdial2010introduction}) is a ``type of computer programming in which the goal is to create something expressive rather than functional''~\cite{creative_coding}, with visual graphics being a common product.  The traditional approach to teaching introductory programming concepts relies heavily on calculation problems where the only output is text output. In contrast, creative coding presents computer science using more familiar contexts~\cite{creative_coding} and allows students to learn programming in a media-rich way. Using programming, students create games \cite{bayliss2006games}, music~\cite{beck2011computing}, stories~\cite{burke2010programming}, physics-based simulations, and image filters~\cite{creative_coding}. Creative tasks such as these make programming more interesting, relevant, and engaging~\cite{greenberg2012creative, tinapple2013digital, xu2018updating}, which in turn improves retention~\cite{tew2005tracking}, motivates students to learn programming~\cite{kelleher2007storytelling}, and broadens participation and diversity in computing by appealing to students who enjoy creative activities and find the traditional programming focus on math and engineering problems overwhelming and/or difficult to relate to~\cite{wolz2011computational, tinapple2013digital, wood2016computational}.

Storytelling, in particular, has long been a popular task in creative coding~\cite{bayon2003mixed}. Prior work (e.g., Kelleher~\cite{kelleher2007storytelling}, Wolz et al.~\cite{wolz2011computational}) has shown that coupling storytelling with programming on block-based programming platforms (e.g., Scratch~\cite{peppler2005creative, maloney2010scratch} or Storytelling Alice~\cite{kelleher2007storytelling}) can motivate students and help them build more confidence and positive attitudes towards programming. However, they use block-based programming and drag-and-drop code editor and focus primarily on making programming more accessible for children and adolescents (ages 8 to 16)~\cite{maloney2010scratch, scratch2021stat}.

To provide a more novice-friendly environment for media-based creative coding aimed at adults, Reas and Fry~\cite{reas2006processing} introduced \textit{Processing}---a text-based programming language that uses the same language conventions and syntax as Java but allows one to create visual, interactive programs like in Scratch~\cite{reas2006processing}). \textit{p5.js}---a JavaScript library offering the same functionality as \textit{Processing}---was also developed~\cite{mccarthy2015getting}, allowing students to easily create interactive graphical programs on the web while learning the syntax and conventions of text-based programming languages they otherwise would have learned through traditional, text-based calculation problems~\cite{guzdial2010introduction}. 

While creating graphic, interactive stories is now possible with text-based programming libraries (e.g., \textit{Processing} and \textit{p5.js}), no work has examined the use of these languages for storytelling-based programming assignments in post-secondary education contexts. Thus it is unclear how to design such assignments, what challenges exist, and how students perceive them. For instance, in collaborative storytelling, individual authors add part of the story following pre-defined constraints (e.g., theme, actions) in order for the parts (contributed by different authors) to form a coherent story. 
In a text-based, creative coding setting, how do we design storytelling assignments to accommodate such procedures? How do we support creative storytelling in individual and collaborative storytelling assignments? Our experience report presents a first step to addressing these questions as we investigate three types of storytelling---collective, individual, and collaborative---through individual and collaborative assignments. In summary, we contribute:

\begin{itemize}
    \item three storytelling assignments for individual, collaborative, and collective storytelling in creative coding class
    \item mixed-methods analysis of students' submissions, experiences, and assessment of the assignments
    \item discussion of benefits and challenges with storytelling assignments in text-based creative coding setting
\end{itemize}

\begin{figure}[htbp!]
    \centering
    \subfloat[\centering Collective storytelling (A6)]{\centering\includegraphics[trim=0cm 1cm 0cm 2cm, clip=true, width=0.49\textwidth]{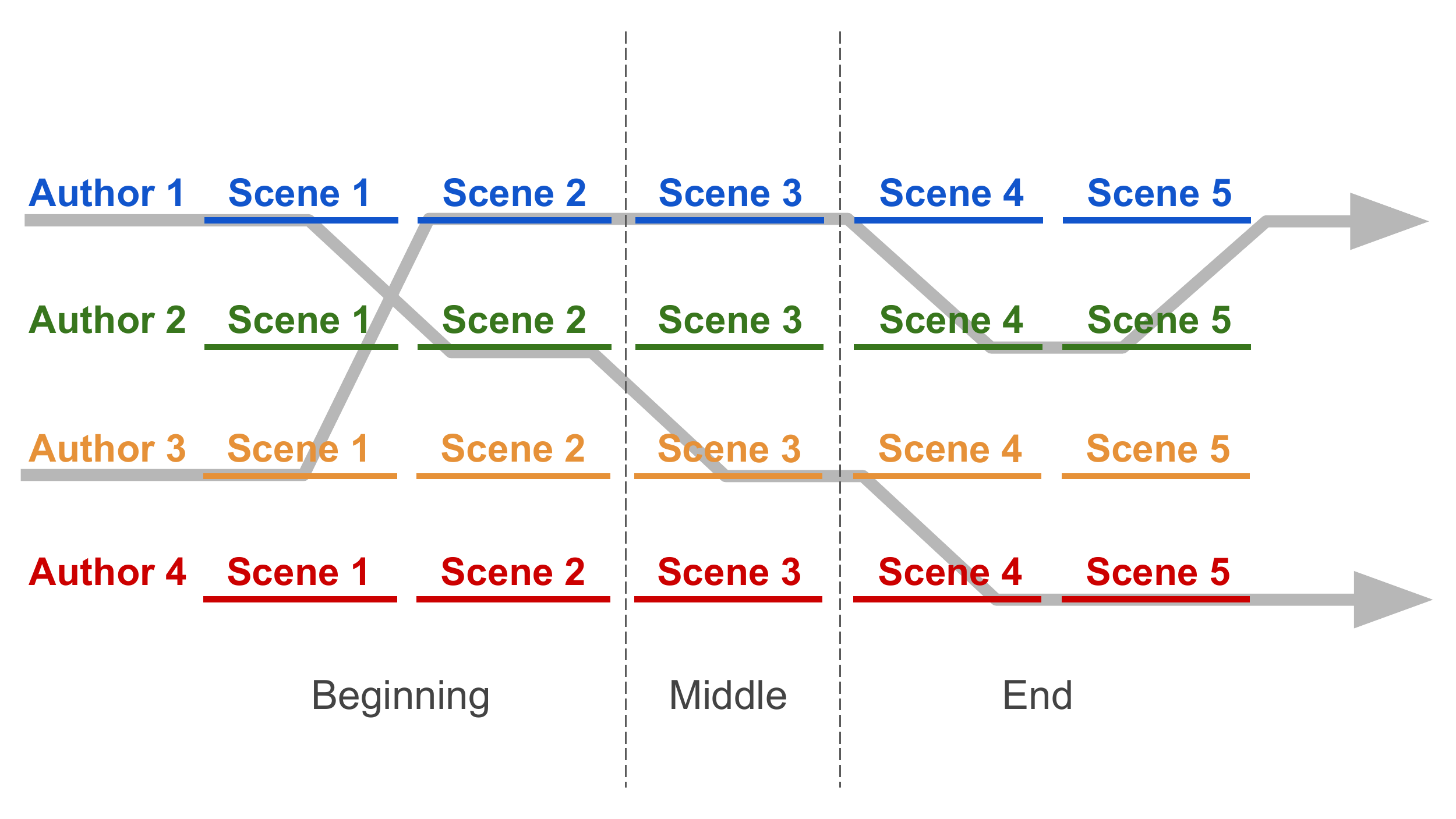} 
    }\\
    \subfloat[\centering Collaborative storytelling (A8)]{\centering\includegraphics[trim=0cm 1cm 0cm 5.5cm, clip=true,width=0.49\textwidth]{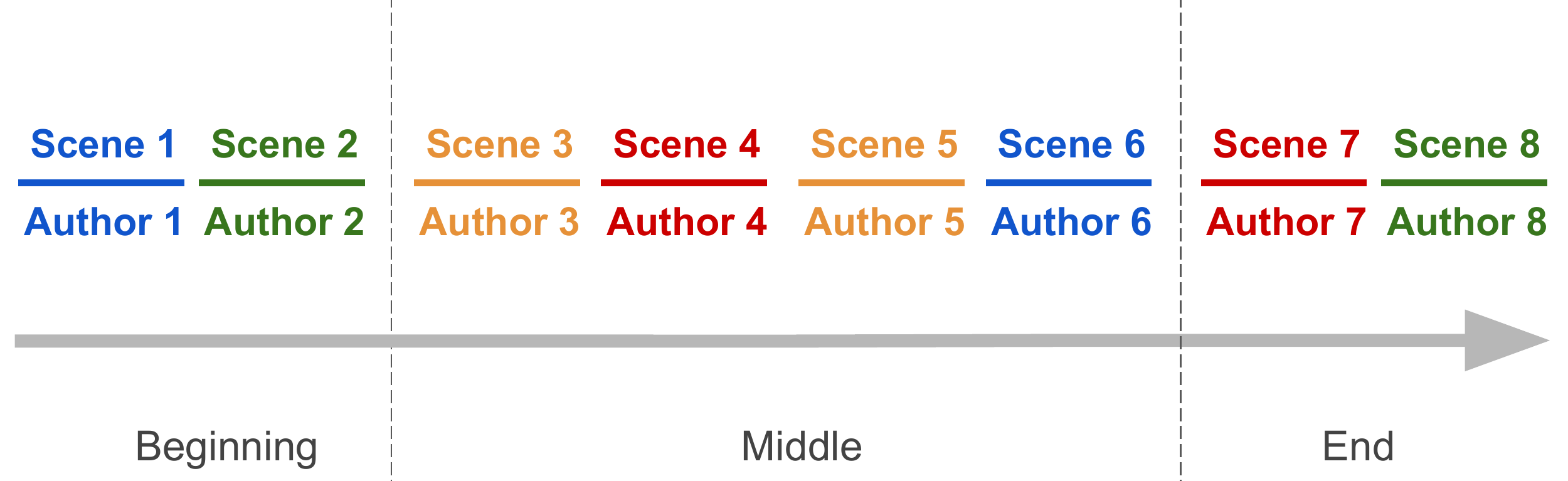} 
    }
    \caption{Authoring procedures we tested for collective (A6) \& collaborative (A8) storytelling. For A6 (top), students write entire stories individually but with design constraints (e.g., theme, number of scenes) so the stories can later be combined and remixed with others' stories, as shown in Fig.~\ref{fig:crowdsourced_narratives}. For A8 (bottom), authors work in relay to add their scenes. Arrows represent story lines readers can explore. As shown, A6 allows multiple story lines (cf. Fig.~\ref{fig:crowdsourced_narratives}), A8 has only one.}
    \label{fig:nonlinear_collaborative_storytelling}
\end{figure}

\begin{figure*}[htbp!]
\centering 
\subfloat[\centering Pre-branch]{\centering\includegraphics[width=0.19\textwidth]{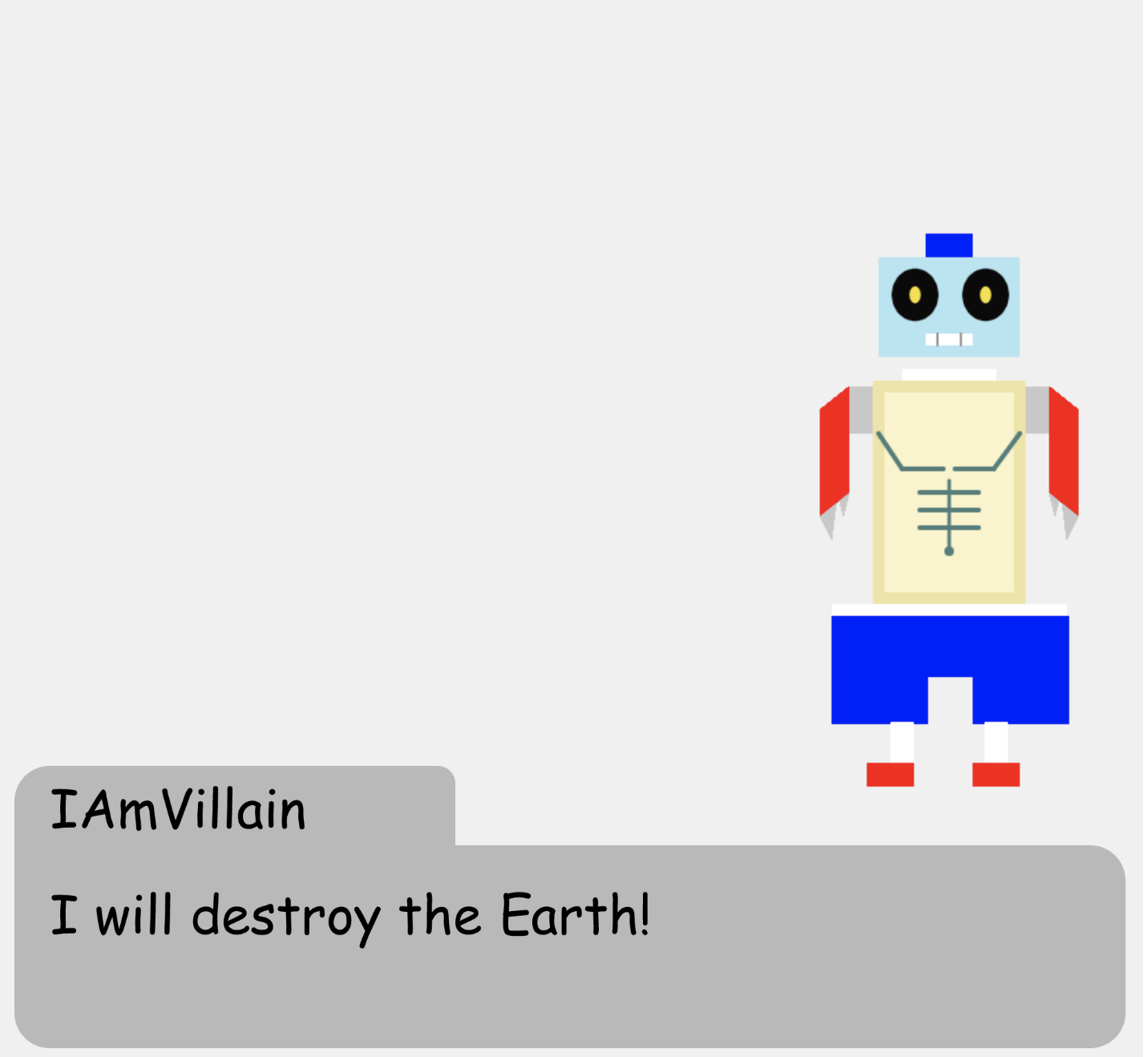}\label{fig:scene1}
}
\subfloat[\centering Pre-branch]{\centering\includegraphics[width=0.19\textwidth]{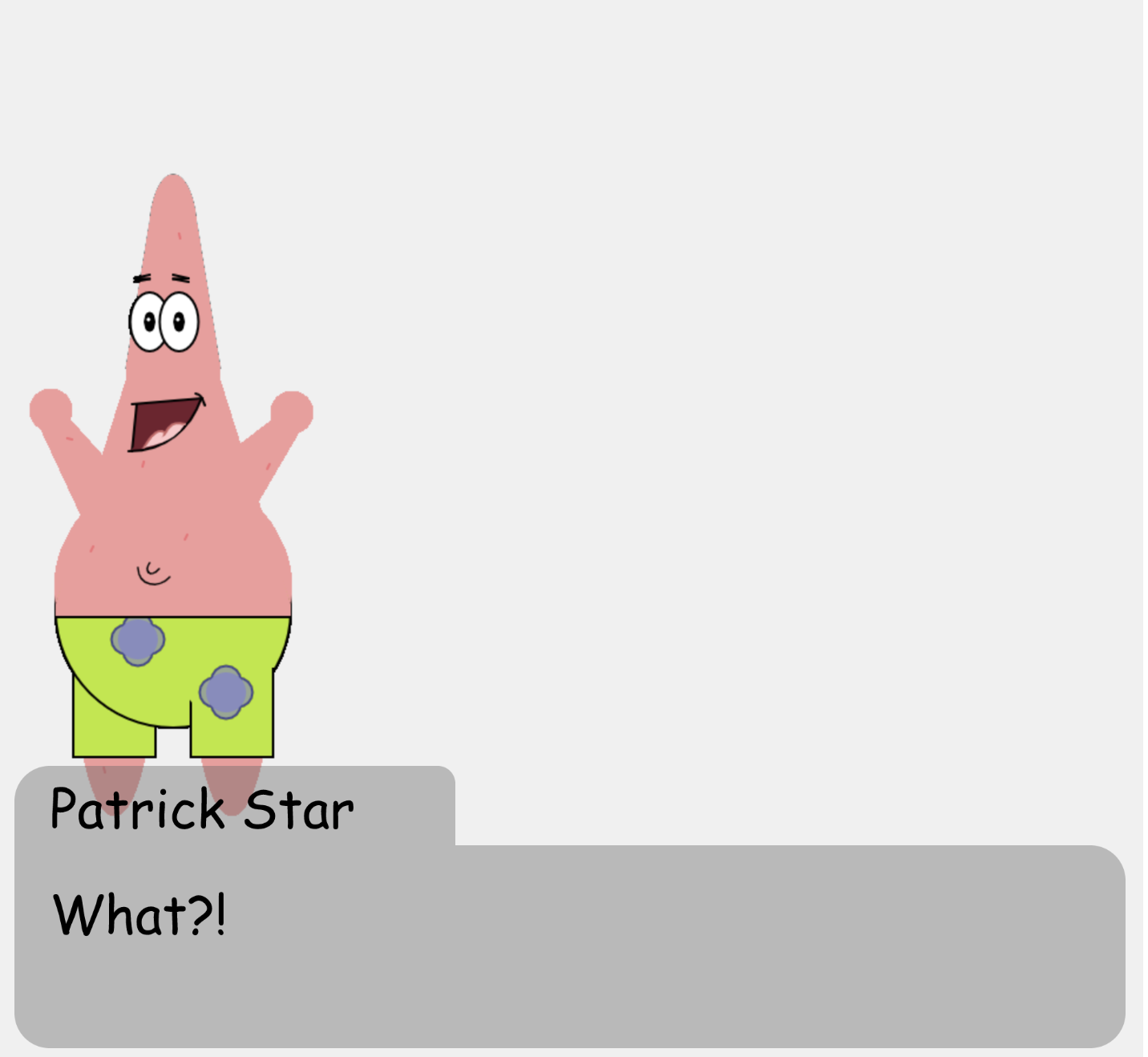}\label{fig:scene2}
}
\subfloat[\centering Branch]{\centering\includegraphics[width=0.19\textwidth]{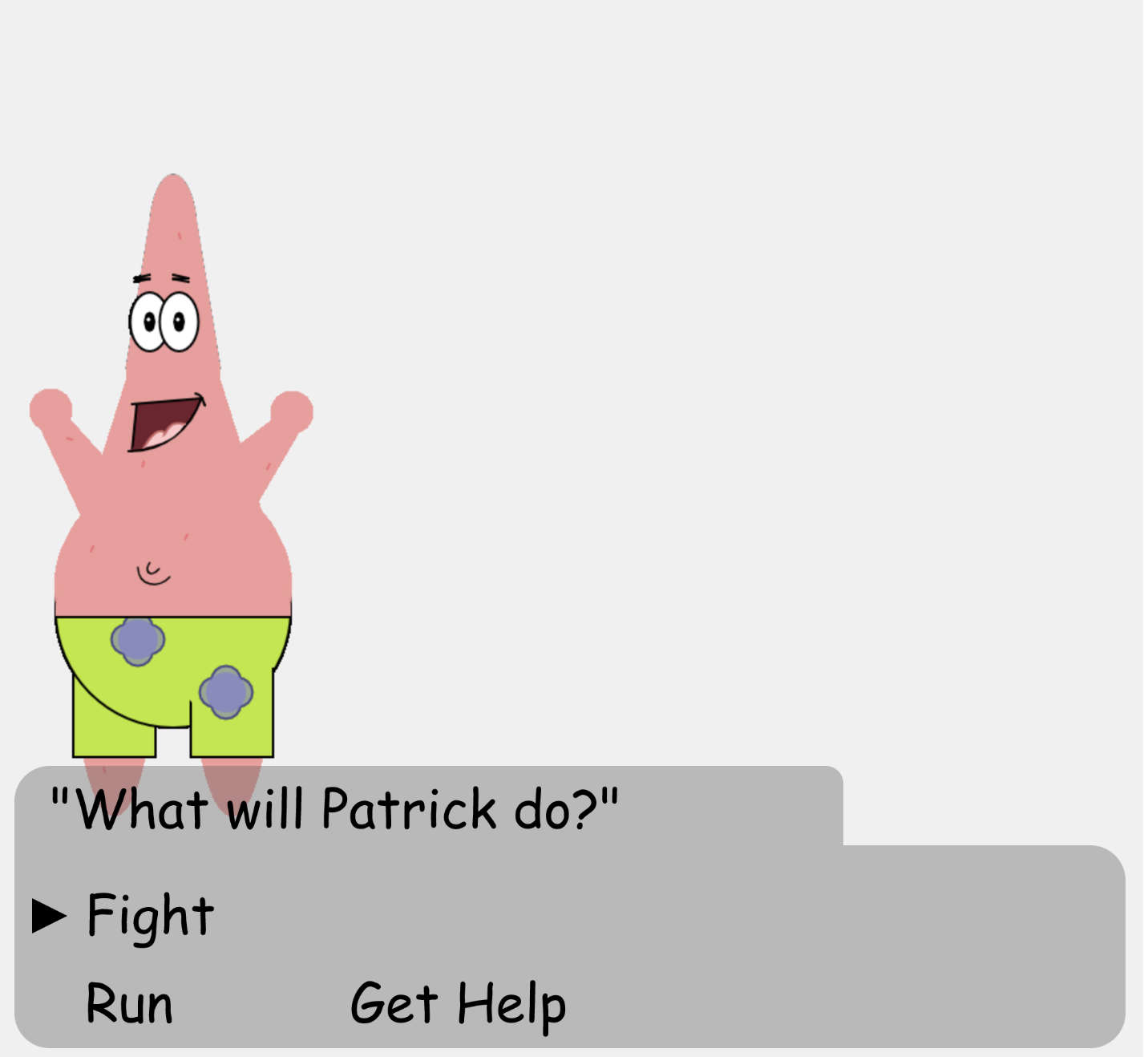}\label{fig:scene3}
}
\subfloat[\centering Post-branch]{\centering\includegraphics[width=0.19\textwidth]{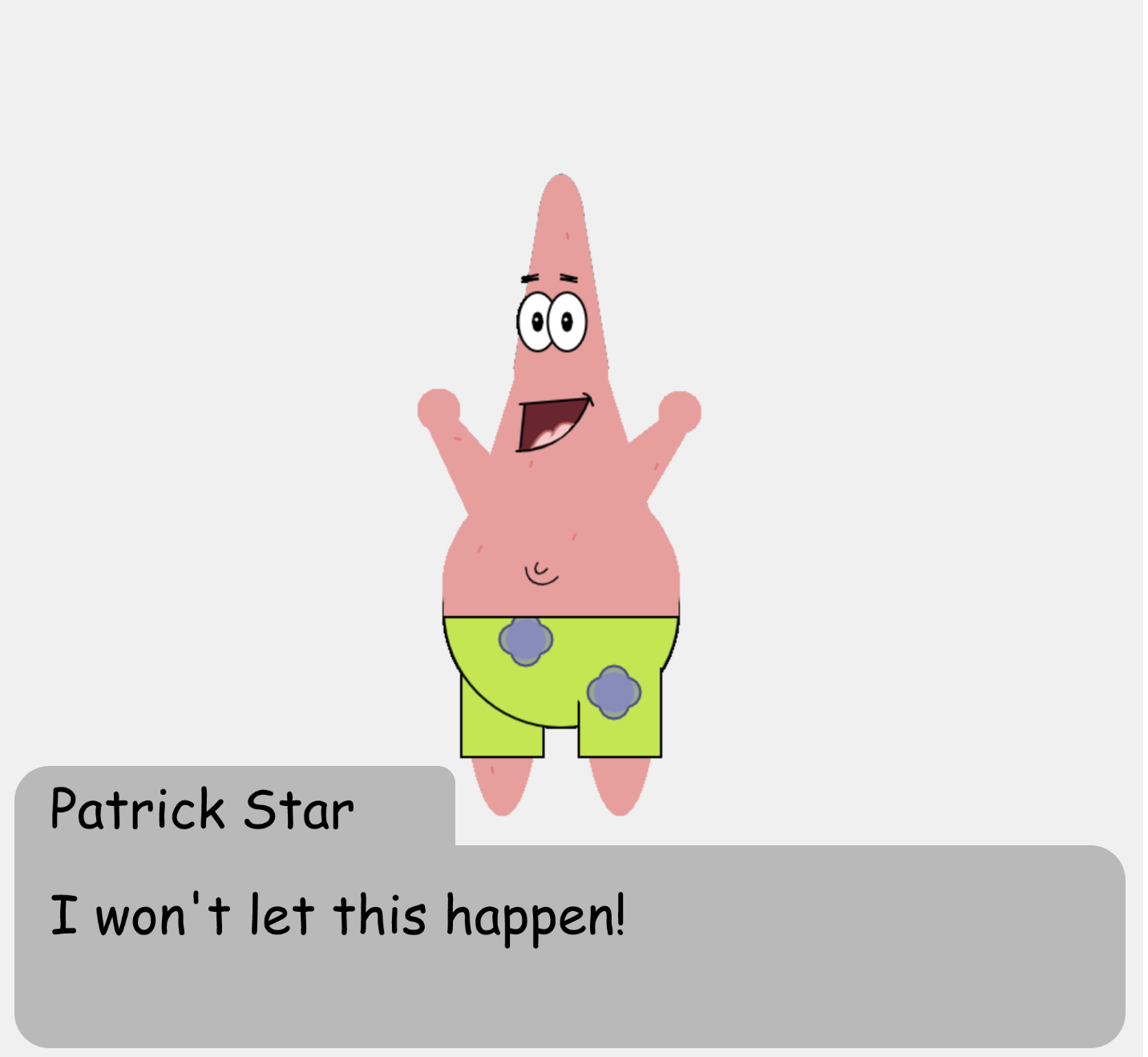}\label{fig:scene4}
}
\subfloat[\centering Post-branch]{\centering\includegraphics[width=0.19\textwidth]{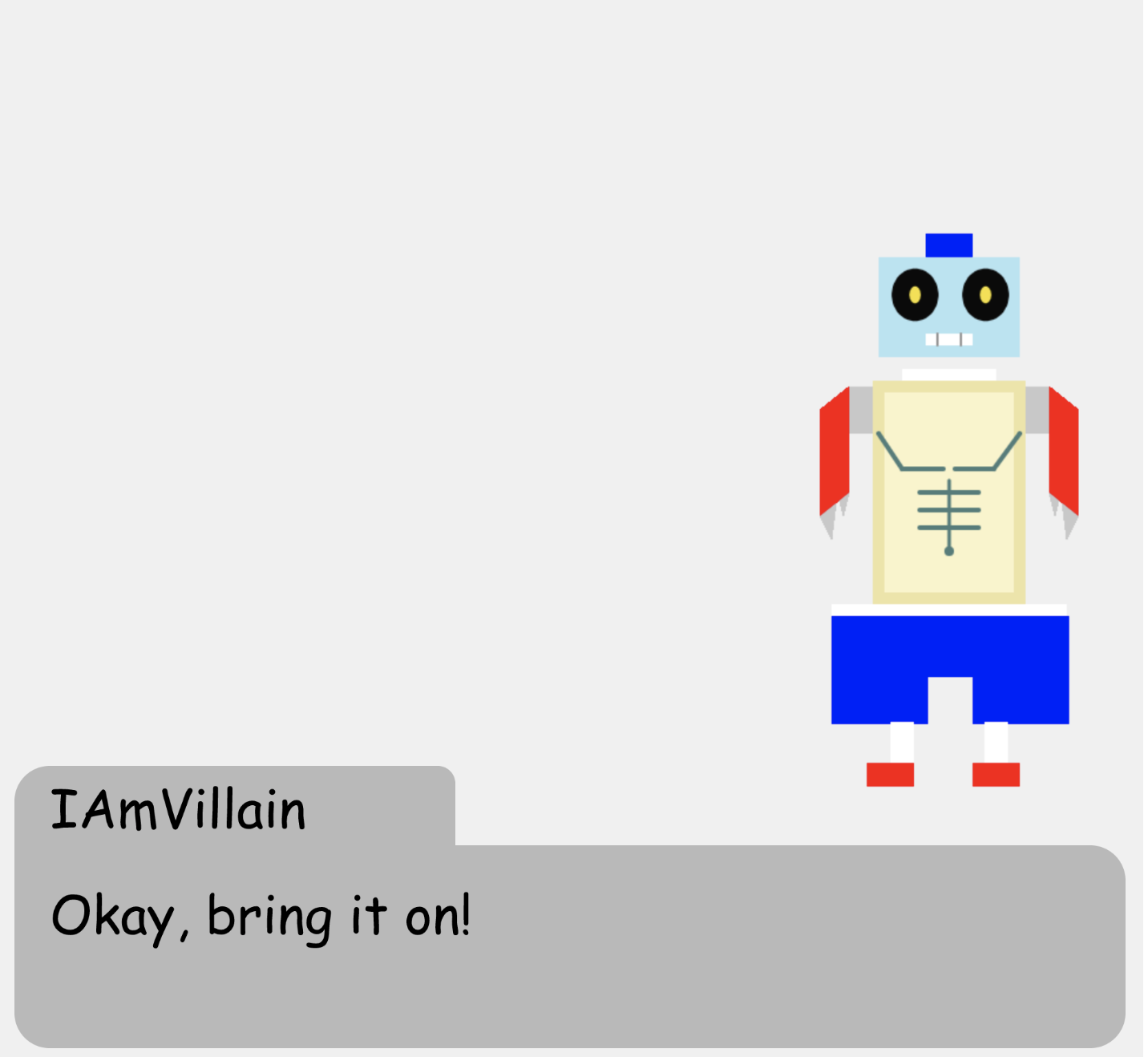}\label{fig:scene5}
}
\caption{Five-scene interactive narrative template--(a) to (e)--on which students were asked to add their own story, characters, background, conflict, and actions. The story progresses from left-to-right, with (c) representing the branching point. For A6, students were asked to come up with their own scenes and dialogues based on pre-defined options (Fight, Run, or Get help) at branch scene (c). For A7 and A8, students were not required to use these options and could define options themselves.}
\label{fig:five_scene_template}
\end{figure*}

\section{Case Study}
We conducted a case study at an introductory computer science course at an R1 university in North America, where students were given individual and collaborative storytelling programming assignments. Below, we describe the course setting, storytelling assignments and how they are designed, our data collection instruments and data analysis methods, and results.

\textbf{Course \& Participants.} At <Anonymized> University, students in the Digital Arts Program are required to take two introduction to computer science courses (I \& II). While open to any non-CS student, these courses are designed primarily for Arts students. Both courses employ a creative coding approach, teaching students the fundamentals of computer programming by having them create interactive graphics such as games and animations with \textit{p5.js}. This report is based on a study administered to students taking the first of the two courses. In the course of 13-weeks, students had eight assignments, which counted for 24\% of the final course grade. The storytelling assignments were administered as the last three assignments (A6, A7 and A8) in the course, beginning in week 6.

Among a total of 49 students who took the course, 42 students answered our post-course anonymous survey and 41 students (15 M, 26 F; no student chose Non-binary) permitted the use and analysis of their survey responses, labs and assignment submissions, and grades for research. Most of the students were in Arts (28), and the rest were in Science (9) and Health Science (4). Most students were undergraduate students in their first and second year (18 First, 15 Second, 4 Third, 3 Fourth, 1 Graduate). In terms of their experience with programming prior to the course, most students had limited experience (15 No Experience, 18 Several Hours/Days, 8 Several Weeks/Months). 

\subsection{Storytelling Assignments}

Our storytelling assignments consisted of two individual (A6 \& A7) and one collaborative (A8) assignments. All storytelling assignments focused on the application of the learned concepts up to that point. At a high level, in all three assignments, students were asked to download a starter project~\footnote{Code available at https://github.com/sanghosuh/storytelling-in-creative-coding} containing the code and image files, update them, and submit them. The main differences between the assignments were the authoring procedure and instructions.

Specifically, when students executed the starter code, they were shown a simple interactive story, shown in Fig.~\ref{fig:five_scene_template}. The story in the starter project featured a ``hero'' structure common in many narratives~\cite{campbell2008hero}, with three characters---protagonist (hero), antagonist (villain), and protagonist's friend. The story has a villain character threatening to destroy the Earth (Fig.~\ref{fig:scene1}-\ref{fig:scene2}), the protagonist (Patrick Star) choosing an action among three options---fight, run, or get help (Fig.~\ref{fig:scene3}), followed by scenes depicting aftermath or reaction to the chosen action (Fig.~\ref{fig:scene4}-\ref{fig:scene5}). As shown in Fig.~\ref{fig:five_scene_template}, the starter project contained three characters, script (i.e., dialogue), specified theme (question \& action choices) and an empty background.

\subsubsection{\textbf{Collective Storytelling (A6)}}
In A6, students were asked to update three components of the story: background, dialogue, and characters. Students were instructed to create three new characters (to replace the characters provided in the starter project): protagonist, antagonist, and protagonist's friend. For this, students were given a link to a website where they could remix exquisite corpse~\cite{exquisite_corpse} submissions (created by students in the class for an earlier assignment in which they were asked to create a character with head, torso, and leg in a specific dimension so the body parts can be remixed with other students' submissions) and download them as image files. There were 42 submissions so they could select one combination from 74,088 possible combinations (42 head * 42 torso * 42 leg). Then they updated a script file in JSON (which specifies the dialogue), added code to set the background, and optionally filled in the empty \texttt{mousePressed} and \texttt{keyPressed} functions to add interactivity to their story~\footnote{\url{https://sanghosuh.github.io/projects/storytelling-in-creative-coding}}.

Students created stories independently, and the instructor combined the stories to form a collective narrative and shared it with students in the following week. Figure~\ref{fig:crowdsourced_narratives} shows a few sample screens from the collective narrative, created by remixing different students' 5-scene interactive narratives. By manipulating the sliders in the top region, readers can change the narrative components: background and interactivity, character, and script. Since there were 21 submissions~\footnote{Students could skip 1 of 8 assignments in this course}, a story could be selected from 9,261 possible combinations (21 backgrounds * 21 characters * 21 scripts). The remixed story was shown before students started working on A7---to interest students in programming and show it as another example of modularity, a concept students learned in weeks prior.

The motivation behind A6 was to administer a collective storytelling model, which allows readers to remix individual stories to explore multiple storylines, as shown in Fig.~\ref{fig:nonlinear_collaborative_storytelling}(a). To enable remix and yet maintain story coherency, we imposed a number of constraints: all stories must use the three action choices [fight, run, get help], branch at scene 3, and have 5 scenes. Fig.~\ref{fig:five_scene_template} shows the general structure of the narrative consisting of 5 scenes with branching occurring at scene 3 and offering 3 possible choices. The story setting is introduced in scenes 1 and 2; the story reaches its peak (in narrative arc) at scene 3 (e.g., What should Patrick do? Decisions: [Fight, Run, Get Help]?); the story shows the aftermath of the peak (release of tension) in scenes 4 and 5~\cite{cohn2013visual}. Having all stories branch at scene 3 based on the same question and a common set of action choices ensured that the remixed story has a consistent theme and remains reasonably coherent (e.g., Fig.~\ref{fig:crowdsourced_narratives}).

\subsubsection{\textbf{Individual Storytelling (A7).}} In A7, as in A6, students were asked to update characters, background, and dialogue in the same manner. Different from A6, the goal of A7 was \textit{not} to remix the stories in the end, but to have students produce individually authored stories without design constraints; students could decide when stories branched, and they could also decide the story theme, i.e., question and decisions (conflict). In other words, students could create stories however they wanted without any design constraints. Students were instructed, however, to create at least 8 scenes and 2 backgrounds. This was to encourage students to be creative and discourage them from submitting stories no different from what they submitted for A6.

\begin{figure}[t!]
\subfloat[\centering Pre-branch scene]{\centering\includegraphics[trim=3.5cm 0.5cm 3.5cm 0.5cm, clip=true, width=.36\textwidth]{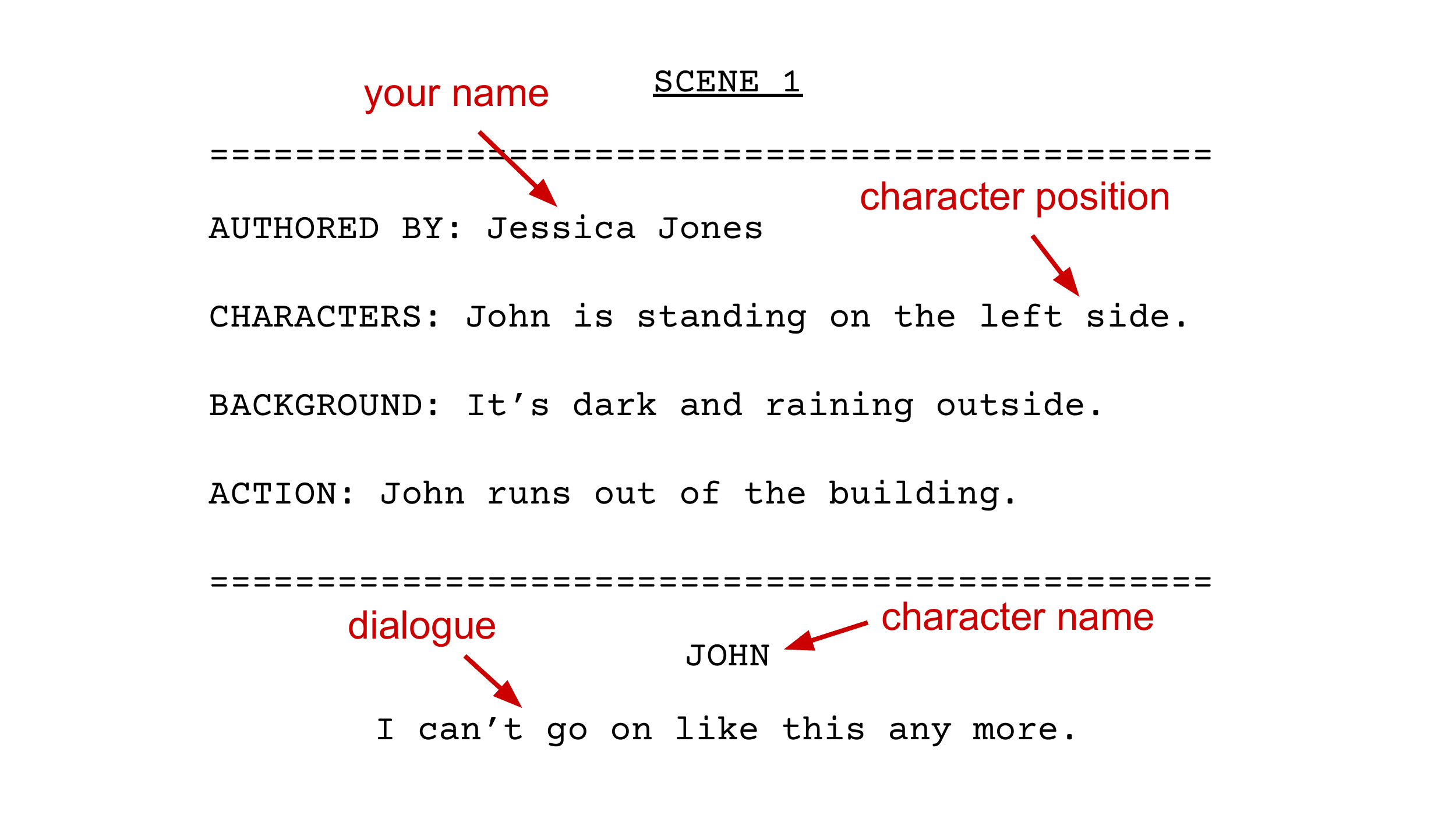} 
}\\
\subfloat[\centering Branch scene]{\centering\includegraphics[trim=3.4cm 0.5cm 3.5cm 0.5cm, clip=true, width=.36\textwidth]{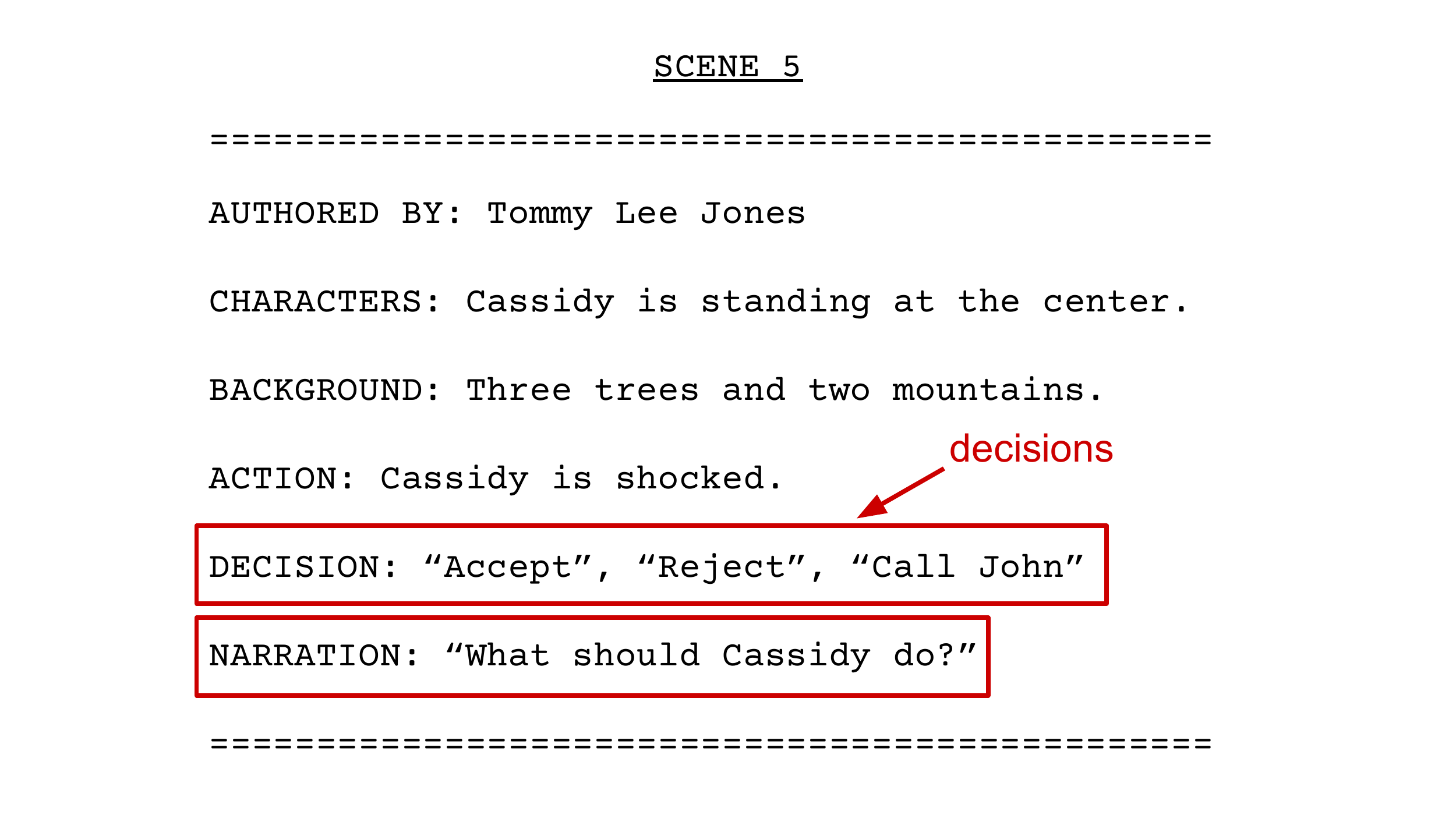}
}\\
\subfloat[\centering Post-branch scene]{\centering\includegraphics[trim=3.5cm 0.5cm 3.5cm 0.5cm, clip=true, width=.36\textwidth]{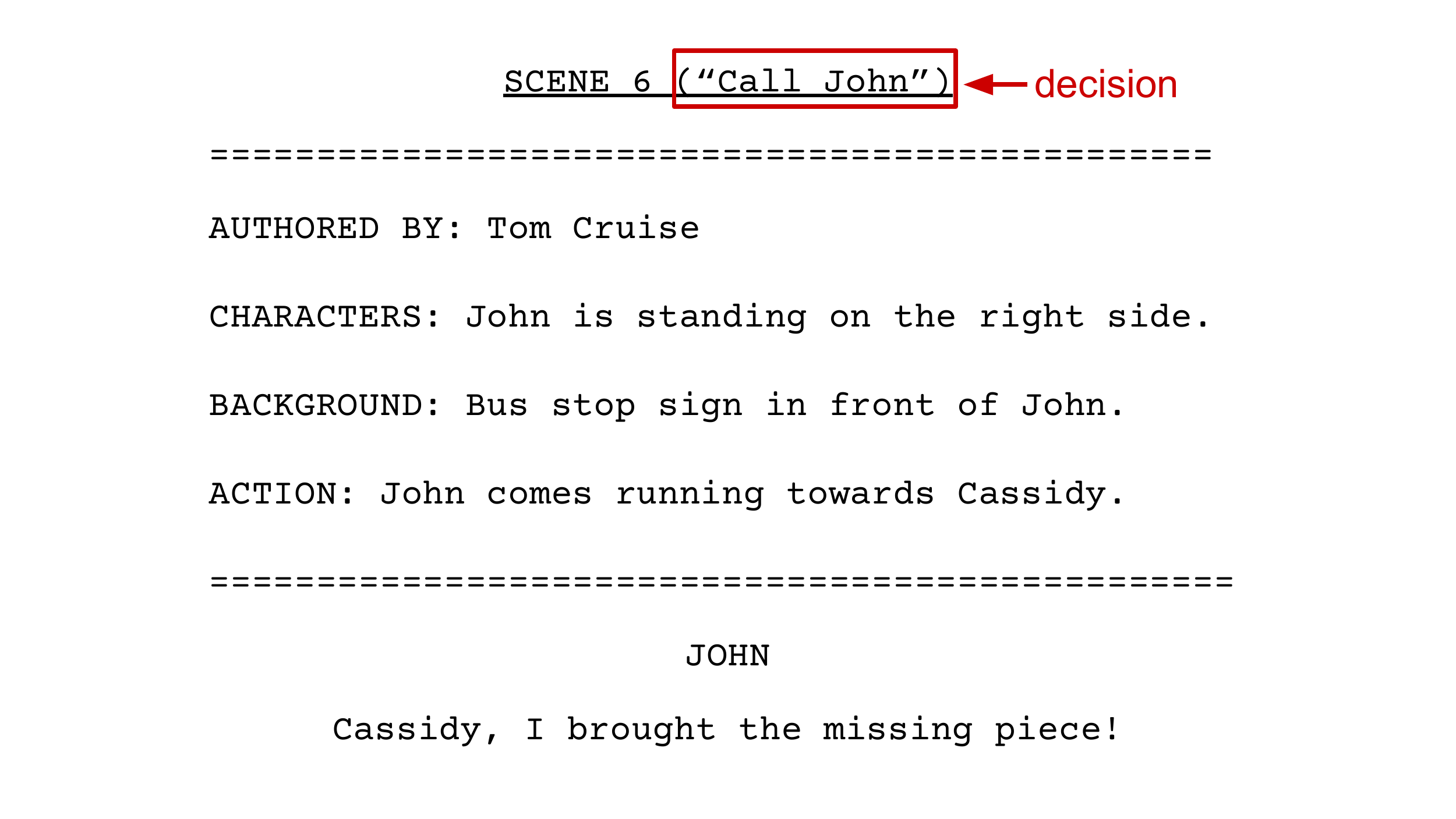}
}
\caption{Assignment 8: Part I - script template for each scene type (pre-branch, branch, post-branch). Students took turns with their group members to add scripts for each scene on Google Docs.}
    \label{fig:a8_part1}
\end{figure}

\subsubsection{\textbf{Collaborative Storytelling (A8).}} In contrast to A6 and A7, A8 involved students working in groups to tell a story in relays.  We divided 48 students (1 student dropped the course before A8) into groups of four or five. This gave us 10 groups---2 groups with 4 students and 8 groups with 5 students. Students worked together as a group to create a story. Due to COVID-19, the class went online after A7 and students were forced to collaborate remotely for A8.

A8 was done in two parts. Part I required students to write a story in English using Google Docs, using the script template shown in Fig.~\ref{fig:a8_part1}. Each student was required to add at least two scenes; the scenes, however, could not be consecutive---to ensure a mix of creative ideas. Each scene contained information about the characters, background, and dialogue, as shown in Fig.~\ref{fig:a8_part1}. 

In the following week, students were asked to implement their scenes for Part II. In Part II, students implemented the story (script) from Part I using a real-time pair programming platform. A small proportion (10\%) of the A8 grade was a group score (given as long as the program did not crash when run by TAs during marking) to motivate students to work with each other and check the project even when finished with their individual scenes. Most (90\%) of the points were individual points to ensure their grades were not unfairly affected by other students' works.


\section{Findings}
Here, we present findings from student submissions for the assignments, as well as a survey administered at the end of the course, to answer how our storytelling assignments supported creative storytelling. The survey had two sets of identical questions---one for individual assignments (A6 and A7) and one for the collaborative assignment (A8). The questions for A6 and A7 were presented together as individual assignments not only to prevent the survey from being too long and repetitive to endanger the response quality, but also because the questions were about their experience with these individual ``assignments,'' not their perceptions nor understanding of collective and individual ``storytelling.'' We are still able to differentiate students' responses between A6 and A7 since we know which assignments each student submitted stories for. To clarify which individual assignment(s) (A6, A7 or both) students likely had in mind when commenting, we tagged onto the subject ID a subscript to indicate which assignments they worked on (e.g., S39$_{6}$, S39$_{7}$, or S39$_{6,7}$). We tagged 0 if a student did not work on any individual storytelling assignment (S39$_{0}$) and only used the feedback pertaining to A8. To ensure anonymity, we refer to students as S1, S2, ... S41 and specify groups in A8 as G1, G2, ... G10.

\subsection{Interactive narratives from individual and collaborative authoring processes}

\subsubsection{Survey Responses.}
To assess whether students were able to use the individual or collaborative storytelling activities to express their creativity, we asked, ``Were you able to be creative in the assignment?'' and ``Did the assignment encourage you to be creative?'' using Likert scales (``Not at all'', ``Slightly'', ``Moderately,'' ``Very,'' ``Extremely'').
Both questions had mostly positive responses (i.e., ``Moderately,'' ``Very,'' or ``Extremely''). For \textit{enabling creativity}: A6 (67\%,
\begin{sparkline}{5} \sparkspike .15 3/21 \sparkspike .317 18/21 \sparkspike .483 15/21 \sparkspike .65 18/21 \sparkspike .817 9/21 \end{sparkline}
), A7 (58\%, \begin{sparkline}{5} \sparkspike .15 12/31 \sparkspike .317 27/31 \sparkspike .483 24/31 \sparkspike .65 15/31 \sparkspike .817 15/31 \end{sparkline}), A8 (60\%, \begin{sparkline}{5} \sparkspike .15 15/40 \sparkspike .317 42/40 \sparkspike .483 27/40 \sparkspike .65 21/40 \sparkspike .817 15/40 \end{sparkline}); \textit{encouraging creativity}: A6 (71\%, \begin{sparkline}{5} \sparkspike .15 6/21 \sparkspike .317 12/21 \sparkspike .483 18/21 \sparkspike .65 24/21 \sparkspike .817 3/21 \end{sparkline}), A7 (65\%, \begin{sparkline}{5} \sparkspike .15 18/31 \sparkspike .317 15/31 \sparkspike .483 24/31 \sparkspike .65 27/31 \sparkspike .817 9/31 \end{sparkline}), A8 (60\%, \begin{sparkline}{5} \sparkspike .15 21/40 \sparkspike .317 27/40 \sparkspike .483 30/40 \sparkspike .65 33/40 \sparkspike .817 9/40 \end{sparkline}).

\subsubsection{Narrative Analysis}

\hfill \break 
\textbf{Sources of Inspiration.} The stories authored individually and collaboratively were both diverse, portraying scenes, characters, and situations from 6 movies (e.g., \textit{Avengers}, \textit{Terminator}), 1 novel (\textit{Charlie and the Chocolate Factory}), 1 TV series (\textit{The Office}), 5 recent events (e.g., movie director Joon-Ho Bong receiving an Oscar), 27 others (original ideas like ``Mom catching Lia skipping classes'' etc.), or 4 based on the Fig.~\ref{fig:five_scene_template} template. 

\begin{figure}[t!]
\centering 
\subfloat[\centering Pokemon]{\centering\includegraphics[width=0.16\textwidth]{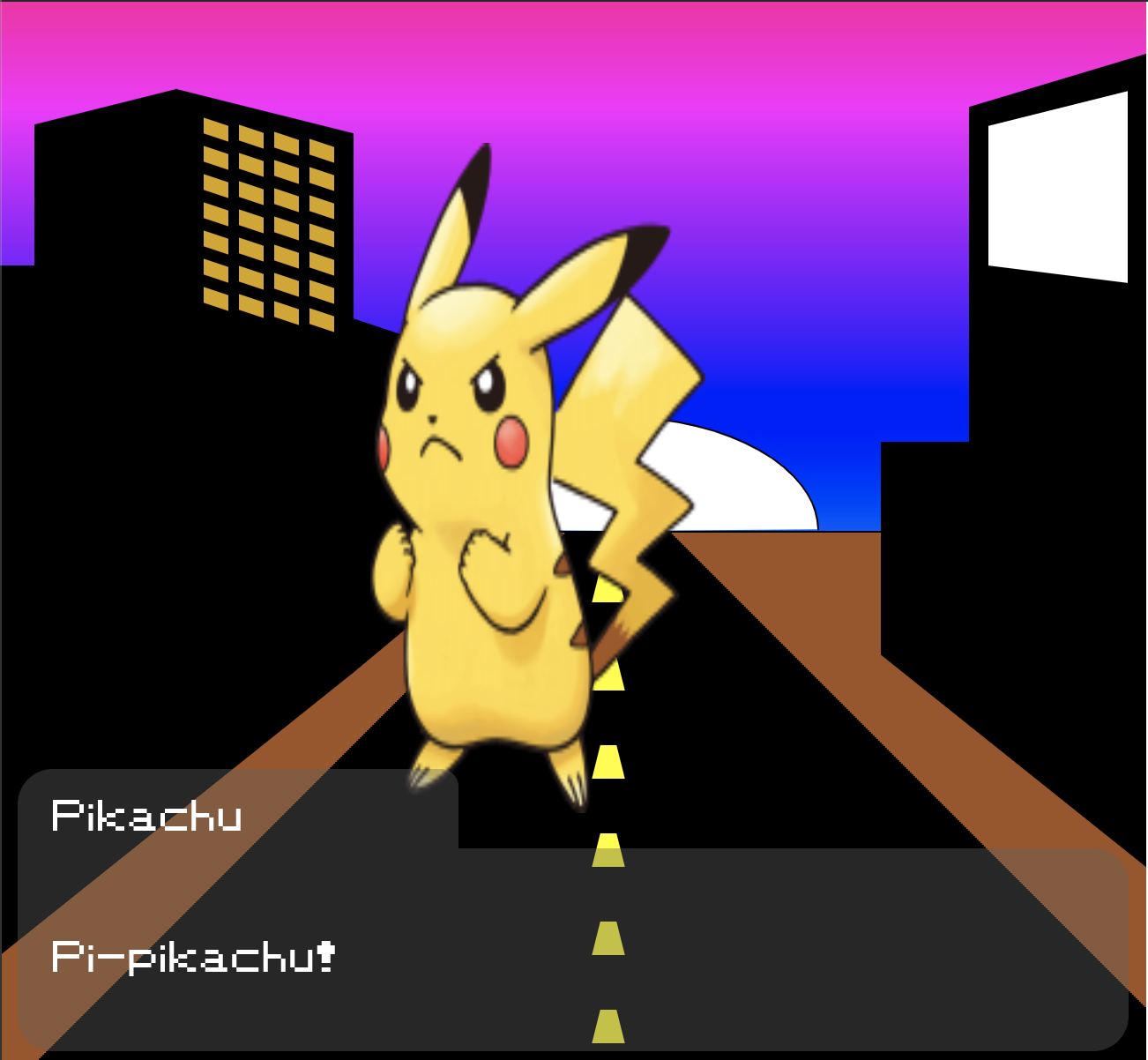}
}
\subfloat[\centering Club]{\centering\includegraphics[width=0.16\textwidth]{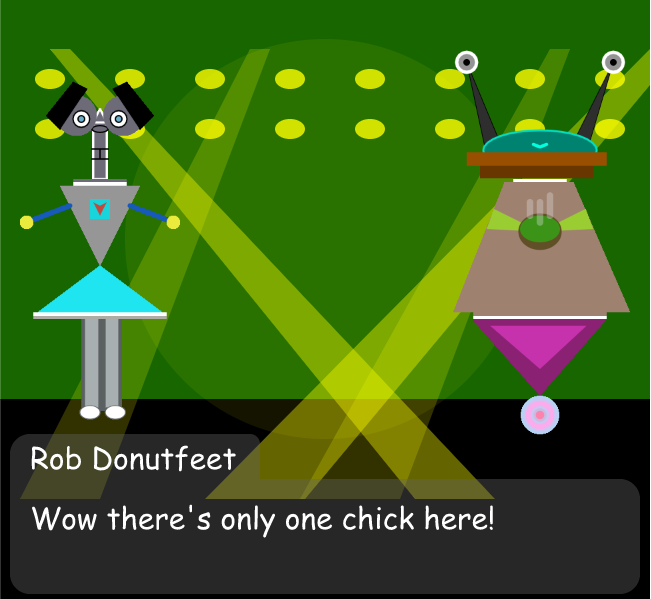}
}
\subfloat[\centering Nemo]{\centering\includegraphics[width=0.16\textwidth]{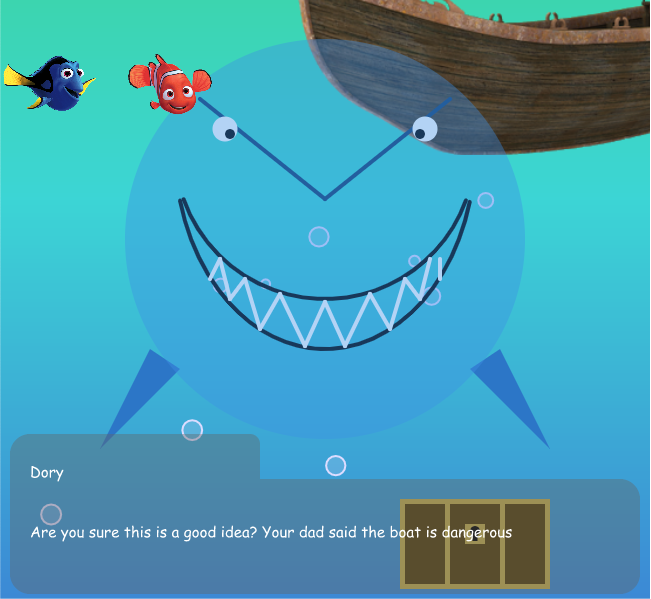}
}
\caption{Examples of background scenes from three storytelling assignments: A6: (a), A7: (b), A8: (c). The scenes are interactive and have animation running in the background.}
\label{fig:background}
\end{figure}

\textbf{Interaction \& Background.} Many students worked hard to create aesthetic scenes. Fig.~\ref{fig:background} shows some examples of background scenes students developed. Fig.~\ref{fig:background}(a) shows a city background with Pikachu standing on the road; the student used one-point perspective to add dimension and quality to the scene. Fig.~\ref{fig:background}(b) shows a club scene, with the lights oscillating back and forth like in a typical dance club. Finally, Fig.~\ref{fig:background}(c) shows a scene from the movie \textit{Nemo} in which he chases after the boat; the scene has water bubbles moving around in the scene to accurately portray life underwater. Readers could interact with these scenes by clicking on the screen and moving the mouse to different locations within the scene, which altered the background color and size of the Sun in the Pikachu scene and made the shark appear right behind Nemo and Dory.

Many students went beyond the minimum required work and used programming concepts to add creative interactions and settings. For instance, hovering a mouse over the enemy's torso area disclosing the enemy's weak point; clicking a light switch turning off the light---the scene turns black; placing a UFO (which follows mouse cursor) above a barn opening the door in the barn and pulling cow towards the UFO. Even though incorporating sound files was not a part of the curriculum, a student (S22$_{6,7}$) and G3 asked the instructor to teach them how to add sound files and then added the Pokemon soundtrack for a battle scene with Pokemon characters (Fig.~\ref{fig:background}a) and a school bell sound to enhance their story.

\textbf{Narrative Structures.} Narrative structures were analyzed to examine whether there exists any difference between the individually and collaboratively authored stories. A6 narratives were excluded from the analysis because they all used the same template structure (Fig. \ref{fig:five_scene_template}) for remixes. 
Both individual A7 and collaborative A8 assignments generated stories with interesting narrative structures, but with collaborative A8 assignments producing some structures not seen in A7. For instance, in A7, nine students (out of the 31 total students who submitted) generated narratives where multiple branches combined into a single branch at the end of the narrative. One student's narrative had a loop structure, where one of the branches loops back to the decision scene: when a user chooses ``Run away'' as opposed to ``Fight in the city'' or ``Fight outside the city'' in the decision scene, the story switches back to the decision scene to force the user (hero) to not ``Run away'' but ``Fight'' (to save people). In A8, two groups' stories had unique structures not observed in A7. Their stories had multiple decision scenes in a row (three for G2 and two for G9) so that branches from the previous decision scene converge to another decision scene.




\smallskip
\noindent Overall, the diversity of stories and their structures suggest that both the individual and collaborative authoring of interactive narratives allowed students to be creative and create stories based on their interests. Examples above also demonstrate how authoring interactive narratives motivated some students to put in extra effort. At times, the collaborative assignment (A8) produced more complex stories than the individual assignment (A7). However, in general, narratives produced through both individual and collaborative authoring processes were equally creative, detailed, and well-designed.

\subsection{Students' experiences with the individual vs. collaborative authoring}

\subsubsection{Rating of Overall Experience}

For this analysis, we used only the data of 17 students who completed all 3 assignments for a more accurate assessment and comparison of the individual and collaborative assignments. The analysis revealed that students, on average, found both individual and collaborative assignments fun, interesting, and engaging while rating individual assignments more positively (fun: 71\% vs. 65\%; interesting: 82\% vs. 65\%; engaging: 82\% vs. 59\%). More students reported having spent extra time on the individual assignments (65\%) than on the collaborative assignment (47\%) because it was fun or engaging. Similarly, more students were satisfied with their individually authored interactive stories (82\%) than with the collaboratively authored stories (47\%). A paired T-test revealed that students gave significantly higher ratings when asked how they felt about the individual assignments (M=4.6/7, SD=1.3) as compared to the collaborative assignments (M=3.4/7, SD=1.5), \textit{t}(16)=2.7, \textit{p}=0.01.

While the analysis shows that students enjoyed individual assignments more and preferred to do them if they had to do it again, they pointed out that collaborative storytelling resulted in more interesting stories. That is, even though students preferred the individual assignments over the collaborative assignment in the context of completing them (\begin{sparkline}{1.5} \sparkspike .2 11/17 \sparkspike .8 6/17 \end{sparkline}), they preferred the collaboratively-authored stories as a reader (\begin{sparkline}{1.5} \sparkspike .2 7/17 \sparkspike .8 10/17 \end{sparkline}; i.e. they found those stories more interesting and preferred to read them over the individually-authored stories).

\subsubsection{Key Differences in the Experience with Individual vs Collaborative Storytelling}

The following analysis is based on the rationale that students provided for the two open-ended questions in the survey, namely (1) What did you like/dislike about the assignment? and (2) Would you recommend [individual/collaborative] storytelling assignments in this class in the future terms? 

\textbf{Communication \& Coordination.} Students faced several challenges as they worked on the collaborative assignment (A8). Notably, they (S10$_{6,7}$, S14$_{7}$, S17$_{7}$, S18$_{6,7}$, S40$_{6,7}$) had difficulties communicating and coordinating with other group members (S21$_{6,7}$, S23$_{6}$); for instance, for S10$_{6,7}$, their group members did not communicate with each other and did not do their parts till the very end. For S9$_{6,7}$, their group members did not implement their scenes in Part II, leaving ``the overall project broken and not running smoothly.'' S40$_{6,7}$ also noted lack of ownership in  a group project---a common problem in group work \cite{schultz1997you}, saying ``everyone just thinks that someone else is going to do it''. While students like S26$_{0}$ were lucky to be ``put into a good group who managed their time well and got [their] respective tasks done on time,'' other students like S10$_{6,7}$ and S27$_{6,7}$ were put in teams with students who did not work to a pre-defined schedule and/or did not communicate well. This posed challenges to these students. 

Some students also reported the experience with the platform we used as ``really frustrating'' (S5$_{6,7}$), with similar negative sentiments reported by many students (S5$_{6,7}$, S11$_{7}$, S18$_{6,7}$, S23$_{6}$, S24$_{6,7}$, S26$_{0}$, S33$_{6,7}$, and S36$_{7}$). Primary reasons included issues with live code syncing issues (e.g., some students' code were not saved when multiple students worked together) and the learning curve associated with using a new platform. Combined with frequent errors students encountered when multiple students edited and ran the code at the same time, these negatively affected students' overall experience with the collaborative assignment.

\textbf{Opportunity to Learn from/with Others.} Students also highlighted benefits and positive experiences with the collaborative storytelling. While some students (S31$_{7}$, S38$_{7}$, S39$_{6}$) simply enjoyed the collaborative nature of the assignment and it being a nice ``change from the usual (individual) assignment'' (S15$_{7}$), many students recognized it as a good ``chance for students to work together'' (S31$_{7}$) and learn from others. Particularly, they saw it as an opportunity to learn ``different coding styles'' (S7$_{7}$), ``thought processes'' (S26$_{0}$), how to code and collaborate with others (S8$_{6}$, S29$_{7}$, S40$_{6,7}$), and important ``programming and life skills'' (S19$_{6,7}$). S7$_{7}$ also pointed that their group provided an access point for help; they shared that their group members ``helped each other when [they] were stuck[, which then] motivated [S7$_{7}$] to do the work.'' 

\textbf{Creative Freedom \& Design Constraints.} Interestingly, all three assignments were seen as providing enough creative freedom---even A6 which we saw as imposing high design constraints. Eleven students explicitly praised the creative freedom with the individual assignments. S15$_{7}$ said that ``they give students a chance to show off their abilities in a fun and creative way.'' S5$_{6,7}$ noted that they required him ``to use both the left (logical thinking, coding) and right side (creativity, writing story) of the brains.'' S7$_{7}$ said, ``I had a lot of fun showcasing what I learned and being creative.'' A few students (S10$_{6,7}$, S26$_{0}$, S29$_{6,7}$), on the other hand, did state that the collaborative assignment allowed more creative freedom than A6 and A7. S10$_{6,7}$ said, ``I liked the creativeness of [individual assignments] but disliked the limitations that were placed... I liked that [the collaborative assignment] was less limiting and we could change things around more.''


\section{Discussion}
While our storytelling assignments allowed and motivated students to use text-based programming (\textit{p5.js}) to author creative, interesting interactive narratives, the administration of such assignments comes with challenges and things to consider. To explore three storytelling approaches, we assigned three storytelling assignments. Some students found them repetitive and suggested skipping A6 or A7. Since group chemistry played an important role in collaborative storytelling (A8), team formation needs to be carefully planned or involve students in the process to ensure successful group work and satisfactory experience---in our study, the instructor assigned students to groups to avoid delays from the group formation process. Some students stated that the issues they experienced in A8 such as communication and coordination would not have occurred if it was done in-person, since they would have worked together in the lab. Thus, we expect collaborative storytelling to be easier and more fun for students in an in-person setting.

There are many ways the storytelling assignments can be administered and extended. For instance, in terms of the template code provided for the assignments, some students with more coding experience found them limiting and stated their preference to code the assignments from scratch. As such, administering a storytelling assignment in which students are asked to create interactive narratives from scratch can be a way to address such needs and serve creative coding classes with more experienced students. The collaborative storytelling assignment can also be used as a fun exercise for learning about version control practices, as they need to engage in tasks such as requesting and accepting pull requests. Finally, we can explore different ways to combine other creative coding tasks with these storytelling assignments. For instance, the template code and instructions can be modified to help students embed games---which they could have created for a previous assignment---into their interactive narratives. For instance, a fighting game a student created can be added so that readers can play them when they choose ``Fight'' in the decision scene.

\section{Conclusion}
 
This work is the first to explore, in the context of postsecondary education, individual and collaborative authoring processes for creating interactive narratives using text-based programming. Our work highlights the benefits and challenges of individual and collaborative approaches, and examines how the two main authoring approaches impacted students' experiences and creativity. Our analysis shows that both individual and collaborative authoring processes allowed students to author creative, interesting interactive narratives based on their experiences, imagination, and existing stories. With more CS courses adopting creative coding curricula, our work contributes an experience report detailing three different storytelling assignments that can be used in creative coding classes. 



\bibliographystyle{ACM-Reference-Format}
\bibliography{main}


\begin{thebibliography}{22}


\ifx \showCODEN    \undefined \def \showCODEN     #1{\unskip}     \fi
\ifx \showDOI      \undefined \def \showDOI       #1{#1}\fi
\ifx \showISBNx    \undefined \def \showISBNx     #1{\unskip}     \fi
\ifx \showISBNxiii \undefined \def \showISBNxiii  #1{\unskip}     \fi
\ifx \showISSN     \undefined \def \showISSN      #1{\unskip}     \fi
\ifx \showLCCN     \undefined \def \showLCCN      #1{\unskip}     \fi
\ifx \shownote     \undefined \def \shownote      #1{#1}          \fi
\ifx \showarticletitle \undefined \def \showarticletitle #1{#1}   \fi
\ifx \showURL      \undefined \def \showURL       {\relax}        \fi
\providecommand\bibfield[2]{#2}
\providecommand\bibinfo[2]{#2}
\providecommand\natexlab[1]{#1}
\providecommand\showeprint[2][]{arXiv:#2}

\bibitem[\protect\citeauthoryear{Bayliss and Strout}{Bayliss and
  Strout}{2006}]%
        {bayliss2006games}
\bibfield{author}{\bibinfo{person}{Jessica~D Bayliss} {and}
  \bibinfo{person}{Sean Strout}.} \bibinfo{year}{2006}\natexlab{}.
\newblock \showarticletitle{Games as a" flavor" of CS1}. In
  \bibinfo{booktitle}{\emph{Proceedings of the 37th SIGCSE technical symposium
  on Computer science education}}. \bibinfo{pages}{500--504}.
\newblock


\bibitem[\protect\citeauthoryear{Bayon, Wilson, Stanton, and Boltman}{Bayon
  et~al\mbox{.}}{2003}]%
        {bayon2003mixed}
\bibfield{author}{\bibinfo{person}{Victor Bayon}, \bibinfo{person}{John~R
  Wilson}, \bibinfo{person}{Danae Stanton}, {and} \bibinfo{person}{Angela
  Boltman}.} \bibinfo{year}{2003}\natexlab{}.
\newblock \showarticletitle{Mixed reality storytelling environments}.
\newblock \bibinfo{journal}{\emph{Virtual Reality}} \bibinfo{volume}{7},
  \bibinfo{number}{1} (\bibinfo{year}{2003}), \bibinfo{pages}{54--63}.
\newblock


\bibitem[\protect\citeauthoryear{Beck, Burg, Heines, and Manaris}{Beck
  et~al\mbox{.}}{2011}]%
        {beck2011computing}
\bibfield{author}{\bibinfo{person}{Robert~E Beck}, \bibinfo{person}{Jennifer
  Burg}, \bibinfo{person}{Jesse~M Heines}, {and} \bibinfo{person}{Bill
  Manaris}.} \bibinfo{year}{2011}\natexlab{}.
\newblock \showarticletitle{Computing and music: a spectrum of sound}. In
  \bibinfo{booktitle}{\emph{Proceedings of the 42nd ACM technical symposium on
  Computer science education}}. \bibinfo{pages}{7--8}.
\newblock


\bibitem[\protect\citeauthoryear{Burke and Kafai}{Burke and Kafai}{2010}]%
        {burke2010programming}
\bibfield{author}{\bibinfo{person}{Quinn Burke} {and} \bibinfo{person}{Yasmin~B
  Kafai}.} \bibinfo{year}{2010}\natexlab{}.
\newblock \showarticletitle{Programming \& storytelling: opportunities for
  learning about coding \& composition}. In
  \bibinfo{booktitle}{\emph{Proceedings of the 9th international conference on
  interaction design and children}}. \bibinfo{pages}{348--351}.
\newblock


\bibitem[\protect\citeauthoryear{Campbell}{Campbell}{2008}]%
        {campbell2008hero}
\bibfield{author}{\bibinfo{person}{Joseph Campbell}.}
  \bibinfo{year}{2008}\natexlab{}.
\newblock \bibinfo{booktitle}{\emph{The hero with a thousand faces}}.
  Vol.~\bibinfo{volume}{17}.
\newblock \bibinfo{publisher}{New World Library}.
\newblock


\bibitem[\protect\citeauthoryear{Cohn}{Cohn}{2013}]%
        {cohn2013visual}
\bibfield{author}{\bibinfo{person}{Neil Cohn}.}
  \bibinfo{year}{2013}\natexlab{}.
\newblock \showarticletitle{Visual narrative structure}.
\newblock \bibinfo{journal}{\emph{Cognitive science}} \bibinfo{volume}{37},
  \bibinfo{number}{3} (\bibinfo{year}{2013}), \bibinfo{pages}{413--452}.
\newblock


\bibitem[\protect\citeauthoryear{Greenberg, Kumar, and Xu}{Greenberg
  et~al\mbox{.}}{2012}]%
        {greenberg2012creative}
\bibfield{author}{\bibinfo{person}{Ira Greenberg}, \bibinfo{person}{Deepak
  Kumar}, {and} \bibinfo{person}{Dianna Xu}.} \bibinfo{year}{2012}\natexlab{}.
\newblock \showarticletitle{Creative coding and visual portfolios for CS1}. In
  \bibinfo{booktitle}{\emph{Proceedings of the 43rd ACM technical symposium on
  Computer Science Education}}. \bibinfo{pages}{247--252}.
\newblock


\bibitem[\protect\citeauthoryear{Guzdial and Ericson}{Guzdial and
  Ericson}{2010}]%
        {guzdial2010introduction}
\bibfield{author}{\bibinfo{person}{Mark Guzdial} {and} \bibinfo{person}{Barbara
  Ericson}.} \bibinfo{year}{2010}\natexlab{}.
\newblock \showarticletitle{Introduction to computing and programming in
  Python: a multimedia approach}.
\newblock  (\bibinfo{year}{2010}).
\newblock


\bibitem[\protect\citeauthoryear{Kelleher, Pausch, and Kiesler}{Kelleher
  et~al\mbox{.}}{2007}]%
        {kelleher2007storytelling}
\bibfield{author}{\bibinfo{person}{Caitlin Kelleher}, \bibinfo{person}{Randy
  Pausch}, {and} \bibinfo{person}{Sara Kiesler}.}
  \bibinfo{year}{2007}\natexlab{}.
\newblock \showarticletitle{Storytelling alice motivates middle school girls to
  learn computer programming}. In \bibinfo{booktitle}{\emph{Proceedings of the
  SIGCHI conference on Human factors in computing systems}}.
  \bibinfo{pages}{1455--1464}.
\newblock


\bibitem[\protect\citeauthoryear{Maloney, Resnick, Rusk, Silverman, and
  Eastmond}{Maloney et~al\mbox{.}}{2010}]%
        {maloney2010scratch}
\bibfield{author}{\bibinfo{person}{John Maloney}, \bibinfo{person}{Mitchel
  Resnick}, \bibinfo{person}{Natalie Rusk}, \bibinfo{person}{Brian Silverman},
  {and} \bibinfo{person}{Evelyn Eastmond}.} \bibinfo{year}{2010}\natexlab{}.
\newblock \showarticletitle{The scratch programming language and environment}.
\newblock \bibinfo{journal}{\emph{ACM Transactions on Computing Education
  (TOCE)}} \bibinfo{volume}{10}, \bibinfo{number}{4} (\bibinfo{year}{2010}),
  \bibinfo{pages}{1--15}.
\newblock


\bibitem[\protect\citeauthoryear{McCarthy, Reas, and Fry}{McCarthy
  et~al\mbox{.}}{2015}]%
        {mccarthy2015getting}
\bibfield{author}{\bibinfo{person}{Lauren McCarthy}, \bibinfo{person}{Casey
  Reas}, {and} \bibinfo{person}{Ben Fry}.} \bibinfo{year}{2015}\natexlab{}.
\newblock \bibinfo{booktitle}{\emph{Getting Started with P5. js: Making
  Interactive Graphics in JavaScript and Processing}}.
\newblock \bibinfo{publisher}{Maker Media, Inc.}
\newblock


\bibitem[\protect\citeauthoryear{Peppler and Kafai}{Peppler and Kafai}{2005}]%
        {peppler2005creative}
\bibfield{author}{\bibinfo{person}{K Peppler} {and} \bibinfo{person}{Y Kafai}.}
  \bibinfo{year}{2005}\natexlab{}.
\newblock \showarticletitle{Creative coding: Programming for personal
  expression}.
\newblock \bibinfo{journal}{\emph{Retrieved August}} \bibinfo{volume}{30},
  \bibinfo{number}{2008} (\bibinfo{year}{2005}), \bibinfo{pages}{314}.
\newblock


\bibitem[\protect\citeauthoryear{Reas and Fry}{Reas and Fry}{2006}]%
        {reas2006processing}
\bibfield{author}{\bibinfo{person}{Casey Reas} {and} \bibinfo{person}{Ben
  Fry}.} \bibinfo{year}{2006}\natexlab{}.
\newblock \showarticletitle{Processing Code: Programming within the Context of
  Visual Art and Design}.
\newblock \bibinfo{journal}{\emph{Aesthetic Computing}} (\bibinfo{year}{2006}),
  \bibinfo{pages}{476}.
\newblock


\bibitem[\protect\citeauthoryear{Schultz}{Schultz}{1997}]%
        {schultz1997you}
\bibfield{author}{\bibinfo{person}{Katherine Schultz}.}
  \bibinfo{year}{1997}\natexlab{}.
\newblock \showarticletitle{“Do You Want to Be in My Story?”: Collaborative
  Writing in an Urban Elementary Classroom}.
\newblock \bibinfo{journal}{\emph{Journal of Literacy Research}}
  \bibinfo{volume}{29}, \bibinfo{number}{2} (\bibinfo{year}{1997}),
  \bibinfo{pages}{253--287}.
\newblock


\bibitem[\protect\citeauthoryear{Tew, Fowler, and Guzdial}{Tew
  et~al\mbox{.}}{2005}]%
        {tew2005tracking}
\bibfield{author}{\bibinfo{person}{Allison~Elliott Tew},
  \bibinfo{person}{Charles Fowler}, {and} \bibinfo{person}{Mark Guzdial}.}
  \bibinfo{year}{2005}\natexlab{}.
\newblock \showarticletitle{Tracking an innovation in introductory CS education
  from a research university to a two-year college}.
\newblock \bibinfo{journal}{\emph{ACM SIGCSE Bulletin}} \bibinfo{volume}{37},
  \bibinfo{number}{1} (\bibinfo{year}{2005}), \bibinfo{pages}{416--420}.
\newblock


\bibitem[\protect\citeauthoryear{Tinapple, Sadauskas, and Olson}{Tinapple
  et~al\mbox{.}}{2013}]%
        {tinapple2013digital}
\bibfield{author}{\bibinfo{person}{David Tinapple}, \bibinfo{person}{John
  Sadauskas}, {and} \bibinfo{person}{Loren Olson}.}
  \bibinfo{year}{2013}\natexlab{}.
\newblock \showarticletitle{Digital culture creative classrooms (DC3) teaching
  21st century proficiencies in high schools by engaging students in creative
  digital projects}. In \bibinfo{booktitle}{\emph{Proceedings of the 12th
  International Conference on Interaction Design and Children}}.
  \bibinfo{pages}{380--383}.
\newblock


\bibitem[\protect\citeauthoryear{Unknown}{Unknown}{2020a}]%
        {creative_coding}
\bibfield{author}{\bibinfo{person}{Unknown}.} \bibinfo{year}{2020}\natexlab{a}.
\newblock \bibinfo{title}{Creative Coding}.
\newblock
\newblock
\urldef\tempurl%
\url{https://en.wikipedia.org/wiki/Creative_coding}
\showURL{%
\tempurl}
\newblock
\shownote{Last accessed 14 February 2020.}


\bibitem[\protect\citeauthoryear{Unknown}{Unknown}{2020b}]%
        {exquisite_corpse}
\bibfield{author}{\bibinfo{person}{Unknown}.} \bibinfo{year}{2020}\natexlab{b}.
\newblock \bibinfo{title}{Exquisite Corpse}.
\newblock
\newblock
\urldef\tempurl%
\url{https://en.wikipedia.org/wiki/Exquisite_corpse}
\showURL{%
\tempurl}
\newblock
\shownote{Last accessed 14 February 2020.}


\bibitem[\protect\citeauthoryear{Unknown}{Unknown}{2021}]%
        {scratch2021stat}
\bibfield{author}{\bibinfo{person}{Unknown}.} \bibinfo{year}{2021}\natexlab{}.
\newblock \bibinfo{title}{Scratch Statistics}.
\newblock
\newblock
\urldef\tempurl%
\url{https://scratch.mit.edu/statistics/}
\showURL{%
\tempurl}
\newblock
\shownote{Last accessed 7 January 2021.}


\bibitem[\protect\citeauthoryear{Wolz, Stone, Pearson, Pulimood, and
  Switzer}{Wolz et~al\mbox{.}}{2011}]%
        {wolz2011computational}
\bibfield{author}{\bibinfo{person}{Ursula Wolz}, \bibinfo{person}{Meredith
  Stone}, \bibinfo{person}{Kim Pearson}, \bibinfo{person}{Sarah~Monisha
  Pulimood}, {and} \bibinfo{person}{Mary Switzer}.}
  \bibinfo{year}{2011}\natexlab{}.
\newblock \showarticletitle{Computational thinking and expository writing in
  the middle school}.
\newblock \bibinfo{journal}{\emph{ACM Transactions on Computing Education
  (TOCE)}} \bibinfo{volume}{11}, \bibinfo{number}{2} (\bibinfo{year}{2011}),
  \bibinfo{pages}{1--22}.
\newblock


\bibitem[\protect\citeauthoryear{Wood, Muhl, and Hicks}{Wood
  et~al\mbox{.}}{2016}]%
        {wood2016computational}
\bibfield{author}{\bibinfo{person}{Zoe~J Wood}, \bibinfo{person}{Paul Muhl},
  {and} \bibinfo{person}{Katelyn Hicks}.} \bibinfo{year}{2016}\natexlab{}.
\newblock \showarticletitle{Computational art: Introducing high school students
  to computing via art}. In \bibinfo{booktitle}{\emph{Proceedings of the 47th
  ACM Technical Symposium on Computing Science Education}}.
  \bibinfo{pages}{261--266}.
\newblock


\bibitem[\protect\citeauthoryear{Xu, Wolz, Kumar, and Greenburg}{Xu
  et~al\mbox{.}}{2018}]%
        {xu2018updating}
\bibfield{author}{\bibinfo{person}{Dianna Xu}, \bibinfo{person}{Ursula Wolz},
  \bibinfo{person}{Deepak Kumar}, {and} \bibinfo{person}{Ira Greenburg}.}
  \bibinfo{year}{2018}\natexlab{}.
\newblock \showarticletitle{Updating Introductory Computer Science with
  Creative Computation}. In \bibinfo{booktitle}{\emph{Proceedings of the 49th
  ACM Technical Symposium on Computer Science Education}}.
  \bibinfo{pages}{167--172}.
\newblock


\end{thebibliography}

\end{document}